\documentclass[acmsmall]{acmart}

\setcopyright{cc}
\setcctype{by}
\acmDOI{10.1145/3832104}
\acmYear{2026}
\acmJournal{PACMSE}
\acmVolume{3}
\acmNumber{ISSTA}
\acmArticle{ISSTA013}
\acmMonth{10}
\acmSubmissionID{issta26main-p108-p}
\received{2026-01-24}
\received[accepted]{2026-06-25}

\usepackage{tikz}
\usepackage{amsmath}
\usepackage{xspace}
\usepackage{flushend}
\usepackage{listings}
\usepackage{threeparttable}
\usepackage{xcolor}
\usepackage{lstlinebgrd}
\usepackage{tcolorbox}
\usepackage{booktabs}
\usepackage{soul}
\usepackage{enumitem}
\usepackage[normalem]{ulem}
\usepackage{amsthm}
\usepackage{multirow}

\usepackage{hyperref}
\usepackage{pifont}
\usepackage{adjustbox}

\definecolor{mycolor}{rgb}{0.122, 0.435, 0.698}% Rule colour
\definecolor{gray1}{gray}{0.3}
\definecolor{darkgreen}{rgb}{0.0, 0.5, 0.0}
\definecolor{darkred}{rgb}{0.82, 0.1, 0.26}
\definecolor{shallowgreen}{RGB}{196, 214, 160}
\definecolor{shallowred}{RGB}{217, 149, 143}
\definecolor{darkgreen}{rgb}{0.0, 0.5, 0.0}
\definecolor{darkred}{rgb}{0.82, 0.1, 0.26}
\newcommand{\cmark}{\textcolor{darkgreen}{\ding{51}}}%
\newcommand{\xmark}{\textcolor{darkred}{\ding{55}}}%

\newcommand{\result}[1]{%
\begin{tcolorbox}[colframe=mycolor,boxrule=0.5pt,arc=4pt,
      left=6pt,right=6pt,top=6pt,bottom=6pt,boxsep=0pt,width=\columnwidth]%
      {#1}
\end{tcolorbox}%
}
\newcommand{\sqlancer}{\textsl{SQLancer}\xspace}

\newcommand{\qpg}{\textsl{QPG}\xspace}
\newcommand{\mcg}{\textsl{MCG}\xspace}
\newcommand{\ccg}{\textsl{CCG}\xspace}
\newcommand{\tlp}{\textsl{TLP}\xspace}
\newcommand{\tlpaggregate}{\textsl{$TLP_{a}$}\xspace}
\newcommand{\tlpdistinct}{\textsl{$TLP_{d}$}\xspace}
\newcommand{\tlpwhere}{\textsl{$TLP_{w}$}\xspace}
\newcommand{\tlpgroupby}{\textsl{$TLP_{g}$}\xspace}
\newcommand{\tlphaving}{\textsl{$TLP_{h}$}\xspace}
\newcommand{\hirgen}{\textsl{HirGen}\xspace}
\newcommand{\hirgenoptimization}{\textsl{$HirGen_{opt}$}\xspace}
\newcommand{\hirgenmutation}{\textsl{$HirGen_{mut}$}\xspace}
\newcommand{\emi}{\textsl{EMI}\xspace}
\newcommand{\yinyang}{\textsl{YinYang}\xspace}
\newcommand{\yinyangsat}{\textsl{$YinYang_{sat}$}\xspace}
\newcommand{\yinyangunsat}{\textsl{$YinYang_{unsat}$}\xspace}
\newcommand{\edefuzz}{\textsl{EDEFuzz}\xspace}

\newcommand{\rti}{\textsl{RTI}\xspace}
\newcommand{\sae}{\textsl{SAE}\xspace}
\newcommand{\metaod}{\textsl{MetaOD}\xspace}
\newcommand{\mrop}{\textsl{MROP}\xspace}
\newcommand{\norec}{\textsl{NoREC}\xspace}

\newcommand{\method}{\textsl{MC}\xspace}
\newcommand{\gcov}{\textsl{gcov}\xspace}
\newcommand{\gcovr}{\textsl{gcovr}\xspace}
\newcommand{\mull}{\textsl{Mull}\xspace}
\newcommand{\pcc}{\textsl{PCC}\xspace}

\newcommand{\etal}{\textit{et al}.\xspace}
\newcommand{\ie}{\textit{i}.\textit{e}.\xspace}
\newcommand{\eg}{\textit{e}.\textit{g}.\xspace}

\newcommand{\todo}[1]{}
\renewcommand{\todo}[1]{{\color{red} TODO: {#1}}}
\newcommand{\add}[1]{}
\renewcommand{\add}[1]{{\color{blue} {#1}}}
\newcommand{\del}[1]{}
\renewcommand{\del}[1]{{}}

\newcommand{\bugsymbol}[0]{\includegraphics[width=0.2cm]{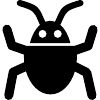}}

\begin{document}

\title{Metamorphic Coverage}

\author{Jinsheng Ba}
\orcid{0000-0003-4008-9225}
\affiliation{%
  \institution{The Chinese University of Hong Kong, Shenzhen}
  \city{Shenzhen}
  \country{China}
}

\author{Yuancheng Jiang}
\orcid{0009-0006-7833-5208}
\affiliation{%
  \institution{National University of Singapore}
  \city{Singapore}
  \country{Singapore}
}

\author{Manuel Rigger}
\orcid{0000-0001-8303-2099}
\affiliation{%
  \institution{National University of Singapore}
  \city{Singapore}
  \country{Singapore}
}

\begin{abstract}
Metamorphic testing is a widely used methodology that examines an expected relation between pairs of executions to automatically find bugs, such as correctness bugs. We found that code coverage cannot accurately measure the extent to which code is validated and mutation testing is computationally expensive for evaluating metamorphic testing methods. In this work, we propose Metamorphic Coverage (MC), a coverage metric that examines the distinct code executed by pairs of test inputs within metamorphic testing. Our intuition is that, typically, a bug can be observed if the corresponding code is executed when executing either test input but not the other one, so covering more differential code covered by pairs of test inputs might be more likely to expose bugs. While most metamorphic testing methods have been based on this general intuition, our work defines and systematically evaluates MC on five widely used metamorphic testing methods for testing database engines, compilers, and constraint solvers. The code measured by MC overlaps with the bug-fix locations of 50 of 64 bugs found by metamorphic testing methods, and MC has a stronger positive correlation with bug numbers than line coverage. MC is 4x more sensitive than line coverage in distinguishing testing methods' effectiveness, and the average value of MC is 6x smaller than line coverage while still capturing the part of the program that is being tested. MC required 359x less time than mutation testing. Based on a case study for an automated database system testing approach, we demonstrate that when used for feedback guidance, MC significantly outperforms code coverage, by finding 41\% more bugs. Consequently, this work might have broad applications for assessing metamorphic testing methods and improving test-case generation.

\end{abstract}




\maketitle

\section{Introduction}

Testing identifies bugs during software development and evolution, and it typically accounts for half of the development expenses~\cite{hailpern2002software}. To automate testing, multiple methods have been proposed that automatically generate or mutate inputs~\cite{li2018fuzzing, godefroid2008grammar}, on which a so-called \emph{test oracle} is applied to determine whether the program's execution behavior is expected~\cite{barr2014oracle}. 

Metamorphic testing is a popular methodology to tackle the test oracle problem~\cite{chen2002metamorphic, segura2018metamorphic}.
It has been applied successfully in various domains, such as database systems~\cite{Rigger2020TLP, Rigger2020NoREC}, compilers~\cite{le2014compiler, le2015finding, le2015randomized}, and Satisfiability Modulo Theory (SMT) solvers~\cite{winterer2020unusual, winterer2020validating, park2021generative}. %, and Web services~\cite{pan2023detecting}.
Metamorphic testing has been proposed as a methodology to test \emph{untestable} systems, that is, systems for which it is difficult to specify the exact behavior that is expected.
Rather than providing a concrete expected output for a given input, metamorphic testing uses a test input $t_1$ to derive one or multiple inputs (\eg, $t_2, \ldots, t_n$), for which a test oracle can be provided that checks whether their outputs 
(\ie, $O_1, \ldots, O_n$)
comply with a relation, which is called \emph{Metamorphic Relation}.
While metamorphic testing can be used to validate non-functional properties such as performance~\cite{cert} or information leakage~\cite{pan2024edefuzz}, we focus on correctness in this paper.

When designing metamorphic testing methods, it is crucial to be able to assess their effectiveness, especially for the researchers who develop such methods. According to our study on ten representative metamorphic testing methods, the most widely used metric to evaluate metamorphic testing methods is the number of found bugs~\cite{chen2018metamorphic, segura2016survey}. However, this metric can only be measured \emph{a posteriori}, that is, after all the bugs found by metamorphic testing have been fixed, as identifying unique bug-inducing test cases for incorrect-output bugs is an open problem~\cite{qian2023survey, yang2023silent}.
This is an issue for researchers, who might want to gauge the potential of a metamorphic relation, before conducting a large-scale testing campaign, which often spans over multiple months~\cite{ding2021empirical}.
Additionally, the quality of target programs can affect the number of unique bugs as a metric because an effective testing method cannot find many bugs in a well-tested program. 
Existing \emph{a priori} metrics, which can gauge the potential of a method before conducting a large-scale testing campaign, have been sparsely adopted.
Code coverage is not often used, presumably because it more precisely captures how effective a test input generator is. Mutation testing, despite advances in improving its efficiency~\cite{papadakis2019mutation}, is still often prohibitively expensive to use in practice.

In this paper, we propose \emph{Metamorphic Coverage} (\method), a simple and practical metric for evaluating the quality of metamorphic testing methods. We believe that a set of test inputs $t_1, t_2, \ldots, t_n$ is typically most effective in finding bugs when the inputs exercise different code paths, since a faulty location might be covered by some inputs, but not the others, resulting in potential violations of the metamorphic relation. Therefore, our idea is to examine the difference in the code exercised by at least one input but not by all inputs to measure the quality of metamorphic relations and metamorphic testing methods.

\method is a coverage metric based on code coverage, which can be any coverage criteria, such as line coverage. Unlike the number of bugs, which is \emph{a posteriori} metric, \method is \emph{a priori} metric that can be measured before conducting a bug-finding campaign. Compared to other coverage metrics, \method can more accurately measure the effectiveness of metamorphic relations. Compared to mutation testing, \method is a lightweight method as it does not require additional execution effort.
\begin{figure}
\begin{lstlisting}[caption={A faulty absolute-difference algorithm implementation.},captionpos=t, label=lst:intro, escapeinside=&&, language=C]
int calculate_difference(int x, int y) {
    if (x > y) {
        return x - y;
    } else {
        return y - x + 1; // &\bugsymbol& y - x;
    }
}
\end{lstlisting}
\end{figure}

\autoref{lst:intro} shows a motivating example. The function calculates the absolute difference between two numbers. We consider two metamorphic relations. $R_1$: if $t_1=(x, y)$ and $t_2=(y, x)$, $O_1=O_2$, indicating swapping both input numbers should output the same result. $R_2$: for an integer $c$, if $t_1=(x, y)$ and $t_2=(x + c, y + c)$, $O_1=O_2$, indicating a constant added to both input numbers should output the same result. For given inputs $x$ and $y$, if $x \neq y$, $t_1$ and $t_2$ of $R_1$ cover the \textbf{\lstinline{if}} and \lstinline{else} branches respectively, while $t_1$ and $t_2$ of $R_2$ cover the same \textbf{\lstinline{if}} or \lstinline{else} branch. Given $t_{1}^\prime=(2,3)$, $t_{1}^{\prime \prime}=(6,2)$ and $c=1$, $R_1$ derives $(3,2)$ and $(2,6)$, and $R_2$ derives $(3,4)$ and $(7,3)$. Both $R_1$ and $R_2$ cover all code lines: $Cov(t_1) \cup Cov(t_2) = \{2-7\}$. However, only $R_1$ can identify the bug in line 5, which is caused by redundant \lstinline{+1}. Thus, line coverage cannot distinguish the bug-finding effectiveness of both relations. Suppose \method is based on line coverage, the code covered by \method is $\method(t_1, t_2) = Cov(t_1) \triangle Cov(t_2) = \{3,5\}$ for $R_1$ and $\varnothing$ for $R_2$. This suggests that $R_1$ has a higher \method and better bug-finding effectiveness than $R_2$, corresponding to our intuition that a relation is more effective in finding bugs when the inputs execute different code paths.

The intuition that the effectiveness of a metamorphic testing method depends on whether the inputs on which the metamorphic relations are defined cover disjoint portions of the code is not new. We believe most metamorphic testing methods are designed based on this intuition~\cite{le2014compiler, ma2023fuzzing, Rigger2020NoREC}. However, this intuition was only informally observed, claimed, or studied on small, artificial metamorphic relations and programs~\cite{cao2013correlation, chen2004case, asrafi2011testing}. We systematically evaluated it on widely used metamorphic testing approaches. We also propose to utilize this intuition in other scenarios, such as guidance-based fuzzing. 

We evaluated \method on five metamorphic testing methods.
The results show that \method is strongly correlated to the bugs found by metamorphic testing methods, since the code covered by \method overlaps with the fixes of 50 of 64 bugs found by the five metamorphic testing methods. \method is 4$\times$ more sensitive than line coverage to distinguish method effectiveness and has the same magnitude of time consumption as line coverage. We used \method as guidance for generating test cases, and it could help the metamorphic testing methods \norec~\cite{Rigger2020NoREC} and \tlp~\cite{Rigger2020TLP} in finding 41\% more bugs than code coverage. This finding has potentially broad implications, enabling efficient feedback-driven test-case generators for metamorphic testing.

Overall, we make the following contributions:
\begin{itemize}
    \item We propose \method, a novel method to evaluate metamorphic testing methods by measuring the differential code coverage of a pair of test inputs.
    \item We implemented and evaluated \method in a comprehensive study on five metamorphic testing methods. The fixes of 50 of 64 bugs found by these methods overlap with the code measured by \method. 
\end{itemize}

\section{Background}\label{sec:background}

\emph{Code coverage.}
Code coverage is the percentage of the source code of a program executed by a particular testing method or test suite. An assumption is that a testing method that covers more code can find more bugs, so code coverage is typically used to assess a testing method's adequacy. Multiple code coverage criteria have been proposed~\cite{gligoric2013comparing}, and in this paper, we consider the following common coverage criteria:
\begin{itemize}
    \item \emph{Line coverage}, which measures the percentage of source code lines that have been executed;
    \item \emph{Statement coverage}, which measures the percentage of instructions that have been executed;
    \item \emph{Branch coverage}, which measures the percentage of control-flow branches (\eg in \lstinline{if} or \lstinline{switch-case} statements) that have been executed.
    \item \emph{Function coverage}, which measures the percentage of functions that have been executed. 
\end{itemize}

To evaluate a metamorphic relation, typically, the cumulative coverage of executing all pairs of $t_a$ and $t_b$ is measured~\cite{winterer2020validating, Rigger2020NoREC}. In \autoref{lst:intro}, line coverage is 100\% (6/6) as all lines have been executed: $\{2-7\}$. Statement coverage is the same as line coverage, because each line only has one statement. Branch coverage is 100\% (2/2), because both branches in lines 3 and 5 have been executed.
The function \lstinline{calculate_difference()} is executed, so the function coverage is 100\% (1/1). Code coverage cannot distinguish both metamorphic relations as both have the same coverage.

\emph{Mutation testing.}
Mutation testing (also known as mutation analysis) is another methodology for assessing the adequacy of a test suite by injecting mutations into programs~\cite{demillo1978hints, jia2010analysis}, and measuring whether the test suite or testing method can identify them as bugs. Mutation testing requires a set of predefined mutation operators $\{m|m \in M\}$, such as mutating an arithmetic operator $-$ to $+$, or removing a function call. We apply $M$ to a program $P$, and obtain a set of program mutations $\{m(P)|m \in M\}$, each of which slightly differs from $P$ and aims to simulate a bug. Given a test suite $X$, if any test $x \in X$ fails when running a $m(P)$, then $m(P)$ is said to be \emph{killed} by $X$, which we denote as $kills(m(P), x)$. Otherwise, $m(P)$ is said to \emph{survive} for $X$. We expect that $X$ can kill more of $\{m(P)|m \in M\}$, so the adequacy of the test suite $X$ is defined by the mutation score, which is computed as the fraction of program mutations killed: $\frac{|\{m(P)|m \in M \& \exists x \in X:kills(m(P), x)\}|}{|\{m(P)|m \in M\}|}$. To evaluate the quality of metamorphic relations, mutation testing can be used to examine how many simulated bugs can be identified as bugs by the metamorphic relations. In \autoref{lst:intro}, suppose the bug in line 5 is a mutation, executing any input violates $R_1$. This mutation is killed and the mutation score is 100\% (1/1). The mutation score is typically deemed a good indication of the fault detection ability of a test suite~\cite{gligoric2013comparing, andrews2005mutation}. However, injecting mutations to simulate bugs is time-intensive~\cite{gorz2023systematic}, because the program needs to be recompiled and executed for each mutant. 
\section{Metamorphic Evaluation Study}\label{sec:study}

As a motivating study, we investigated what metrics were used to evaluate popular metamorphic testing methods.

\emph{Studied metamorphic testing methods.} As shown in \autoref{tab:study}, we chose ten representative metamorphic testing methods. They were published in well-known academic conferences of programming languages and software engineering during the past 10 years (2014--2024). \emph{Referentially Transparent Inputs} (\rti)~\cite{he2021testing} and \emph{Metamorphic Object Detection} (\metaod)~\cite{wang2019metamorphic} test AI systems. \emph{Equivalence Modulo Inputs} (\emi)~\cite{le2014compiler} and \hirgen~\cite{ma2023fuzzing} test the compilation correctness of compilers. \emph{Non-optimizing Reference Engine Construction} (\norec)~\cite{Rigger2020NoREC} and \emph{Ternary Logic Partitioning} (\tlp)~\cite{Rigger2020TLP} test the query processing functionality of Database Management Systems (DBMSs). \yinyang~\cite{winterer2020validating} and \emph{Skeletal Approximation Enumeration} (\sae)~\cite{yao2021skeletal} test Satisfiability Modulo Theory (SMT) solvers, which are used to determine whether there exists an assignment to variables that satisfy a given formula. \emph{Excessive Data Exposure Fuzzing} (\edefuzz)~\cite{pan2023detecting} and \emph{Metamorphic Relation Output Patterns} (\mrop)~\cite{segura2018web} test whether Web applications return complete and correct content. To study how these metamorphic testing methods were evaluated, we carefully examined the papers describing the ten metamorphic testing methods.

\begin{table}
    \centering\small
    \caption{Metrics for evaluating metamorphic testing methods. \emph{Code Coverage} includes line, function, and branch coverage.}
    \begin{tabular}{@{}lllcccc@{}}
    \toprule
            &  & & \multicolumn{4}{c}{\textbf{Metrics}}  \\
       \cmidrule{4-7}
    \textbf{Method} & \textbf{Target} & \textbf{Publication} & \textbf{Bugs} & \textbf{Code Coverage} & \textbf{Mutation Score} & \textbf{Time} \\ 
    \midrule
        \rti~\cite{he2021testing}               & AI            & ICSE'20       & \checkmark &            &            & \checkmark \\ % New bugs, throughput
        \metaod~\cite{wang2019metamorphic}      & AI            & ASE'20        & \checkmark &            &            & \checkmark \\ % New bugs, throughput
        \emi~\cite{le2014compiler}              & Compiler      & PLDI'14       & \checkmark &            &            &            \\ % New bugs
        \hirgen~\cite{ma2023fuzzing}            & Compiler      & ISSTA'23      & \checkmark &            &            &            \\ % New bugs
        \norec~\cite{Rigger2020NoREC}           & DBMS          & OOPSLA'20     & \checkmark &            &            &            \\ % New bugs, bug-finding efficiency
        \tlp~\cite{Rigger2020TLP}               & DBMS          & FSE'20        & \checkmark & \checkmark &            &            \\ % New bugs, bug-finding efficiency, line coverage to investigate the bug overlap quantitatively
        \yinyang~\cite{winterer2020validating}  & SMT           & PLDI'20       & \checkmark & \checkmark &            &            \\ % New bugs, line, function, and branch coverage
        \sae~\cite{yao2021skeletal}             & SMT           & FSE'21        & \checkmark & \checkmark &            & \checkmark \\ % New bugs, line coverage, throughput
        \edefuzz~\cite{pan2023detecting}        & Web           & ICSE'24       & \checkmark &            &            & \checkmark \\ % New bugs, throughput
        \mrop~\cite{segura2018web}              & Web           & ICSE'18       & \checkmark &            & \checkmark & \checkmark \\ % New bugs, throughput, mutation score
    \bottomrule
    \end{tabular}
    \label{tab:study}

\end{table}

\emph{Results.} The column \emph{Metrics} in \autoref{tab:study} shows the four identified evaluation metrics used for evaluating the ten studied metamorphic testing methods. \emph{Bugs} refers to the number of found bugs in the real world. \emph{Time} represents the execution throughput. Overall, \emph{Bugs} is the most prevalent and important evaluation metric as it was used to evaluate all metamorphic testing methods. Three methods \emi, \hirgen, and \norec were evaluated by \emph{Bugs} only. \emph{Code Coverage} and \emph{Mutation Score} are not commonly used, and we believe that the reason for this is the accuracy of code coverage and time cost for mutation testing. For example, concerning code coverage, in \norec~\cite{Rigger2020NoREC}, the authors claimed that \emph{``code coverage is not particularly useful for fuzzing DBMS, since high coverage for the core components (\eg, the query optimizer) can be achieved quickly.''} \tlp and \yinyang measure line coverage, while \sae measures line, function, and branch coverage. Only \mrop reported a mutation score. We suspect that other approaches did not adopt mutation testing due to the high execution time needed to compute it. It shows that only the number of bugs, as a \emph{a posteriori} metric, is widely used, and no \emph{a priori} metric is widely used---we understand an a priori metric as one that eschews an extensive testing campaign that requires developers to fix reported issues to measure the metric's effectiveness. Additionally, we found that half of the studied methods were evaluated by throughput, which shows that testing efficiency is also an important factor for metamorphic testing methods. 

\result{No widely applicable \emph{a priori} metric is used for evaluating metamorphic testing methods.}
\section{Approach}\label{sec:approach}
We propose \emph{Metamorphic Coverage} (\method), a novel metric for evaluating the quality of metamorphic testing methods by measuring the differential code executed by pairs of test inputs. For a pair of test inputs $t=(t_a, t_b)$, our intuition is that a bug can be observed if it affects the execution of either $t_a$ or $t_b$ but not the other, implying that pairs that cover more differential code are more likely to find bugs.

\begin{figure}
    \centering
    \includegraphics[width=0.6\columnwidth]{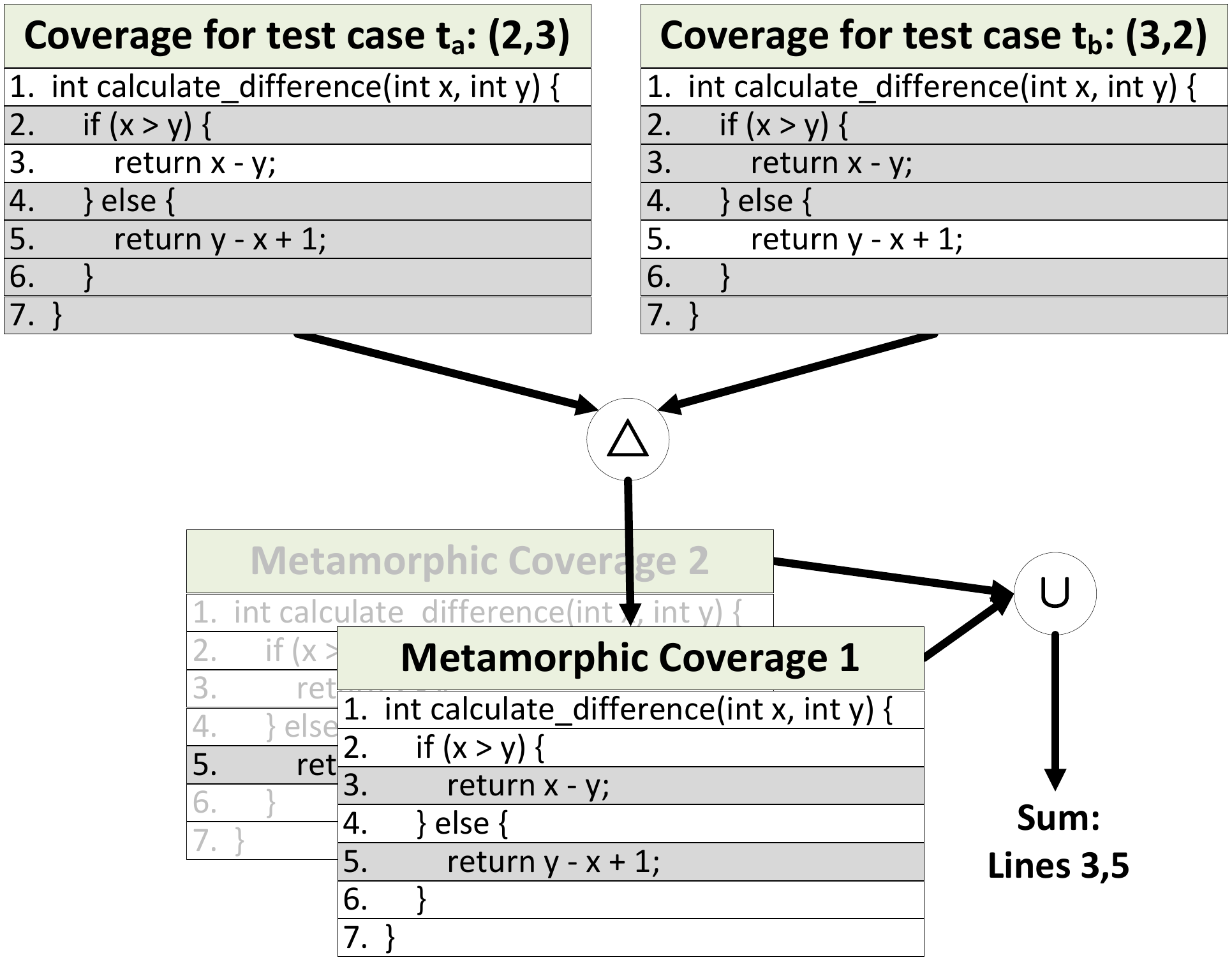}
    \caption{Overview of \method. The gray color above refers to the code covered by executing the program passing $t_a$ and $t_b$, and the gray color below refers to the code covered in differential coverage.}
    \label{fig:overview}
\end{figure}

We define \emph{Metamorphic Coverage (\method)} as follows:
Given a metamorphic relation involving $n$ test inputs $t_1, t_2, \ldots, t_n$, we decompose it into all pairwise combinations $(t_a, t_b)$ for $1 \leq a < b \leq n$.
For an ordered pair of test inputs $t=(t_a, t_b)$, the code covered by metamorphic coverage is $\method(t) = Cov(t_a) \:\triangle\: Cov(t_b) = (Cov(t_a) \cup Cov(t_b)) - (Cov(t_a) \cap Cov(t_b))$, in which $\triangle$ represents the differential coverage and $Cov(t_a)$ and $Cov(t_b)$ represent the code covered by code coverage---any concrete code coverage metric, such as line and branch coverage can be used---of executing $t_a$ and $t_b$ in the target system. Subsequently, for conciseness, we refer to $Cov(t_a) \triangle Cov(t_b)$ as \emph{differential coverage}. Suppose all pairs of test inputs $T=\{t_1, t_2, ..., t_k\}$, the code covered by metamorphic coverage is $\method(T) = \bigcup_{i=1}^{k}(\method(t_i))$. \autoref{fig:overview} illustrates how to compute \method, and we explain the concrete steps to measure \method for each $t$ and to combine them as follows.

\emph{Step 1: collecting coverage.} Given a pair of test inputs $t=(t_a, t_b)$, we first execute the target program passing them as inputs, and measure their executed code for code coverage $Cov(t_a)$ and $Cov(t_b)$ separately. $Cov()$ can be any metric, including line $Cov_{line}()$, branch $Cov_{branch}()$, and function coverage $Cov_{function}()$. \method is not restricted to any specific coverage metrics. In \autoref{fig:overview}, we show the example by $Cov_{line}()$ because it is straightforward to understand. Unless specified, $Cov()$ is short for $Cov_{line}()$. In \autoref{fig:overview}, $Cov(t_a) = \{2, 4, 5, 6, 7\}$, and $Cov(t_b) = \{2, 3, 4, 6, 7\}$.

\emph{Step 2: deriving \method for $t$.} We calculate the code covered by \method by measuring the differential coverage $ Cov(t_a) \:\triangle\: Cov(t_b)$. In \autoref{fig:overview}, this yields Metamorphic Coverage 1: $\method(t) = \{3, 5\}$.

\emph{Step 3: deriving \method for $T$.} Step 2 measures the tested program parts, and we combine all $\method(t)$ by unioning them to derive $\method(T)$ for measuring the tested program parts. In \autoref{fig:overview}, $T$ includes two pairs of test inputs. The \method for second input pair is Metamorphic Coverage 2, $\{5\}$, so $\method(T) = \{3, 5\}$.

\emph{Advantages.}
\method is effective and lightweight in evaluating the quality of metamorphic testing methods. \method considers the code that is validated by metamorphic relations, instead of all executed code. Therefore, \method is more sensitive to distinguishing the quality of different metamorphic testing methods than code coverage. Computing \method does not require recompiling the program, which is necessary for mutation testing, so \method is more lightweight than mutation testing. \method is slightly more expensive than code coverage as computation effort is required to derive differential code coverage.

\emph{Limitations.}
\method is not an absolute indicator of bug-finding effectiveness. 
First, bugs can be found even if \method is zero. Suppose we have a one-line C program \lstinline{int output = input + 1000;}. A possible metamorphic relation is that if an input $t_a$ is bigger than another input $t_b$, $t_a$'s output must be bigger than $t_b$'s output. A bug occurs if $t_a$ is too big so that $t_a+1000$ overflows. However, in this case, \method is zero, as both cases execute the same code. However, in \autoref{subsec:Q1}, we will show that this situation is not prevalent in practice. We have not encountered such a case in our evaluation, and thus believe that this is rare.
Second, an \method score of 100\% does not mean the target program is tested thoroughly. If $t_a$ covers all code, while $t_b$ does not trigger any program logic, \method is 100\%. Typically, we expect executing both inputs to trigger different logic, so that we can compare them to find bugs. If either input does not trigger any logic, no logic is evaluated and the metamorphic relation is too weak.

\emph{Implementation.} We implemented \method for the C/C++ programming language. For step 1, we measured code coverage by \gcov,\footnote{\url{https://gcc.gnu.org/onlinedocs/gcc/Gcov.html}} which is the most widely used source code coverage analysis tool for C/C++. For step 2, we derive the coverage difference by implementing a plugin in \gcovr,\footnote{\url{https://gcovr.com}} which is a popular utility for managing \gcov. \gcovr supports multiple formats to store, analyze, and visualize code coverage. We used the JSON format to store the code coverage collected from step 1, because it is a structured format that facilitates machine processing. In step 2, our plugin derives a new coverage file in the same format of JSON, so that we can leverage \gcovr to visualize and analyze \method without additional implementation effort. The plugin was implemented in around 100 lines of Python code, suggesting that its low implementation effort might make the approach widely applicable.
\section{Evaluation}\label{sec:evaluation}

To evaluate the effectiveness and efficiency of \method for evaluating metamorphic testing methods, we seek to answer the following questions:
\begin{description}
    \item[\textbf{Q.1 Effectiveness.}] Can \method evaluate the bug-finding capability of metamorphic testing methods? 
    \item[\textbf{Q.2 Sensitivity.}] To what extent can \method distinguish the bug-finding capability?
    \item[\textbf{Q.3 Efficiency.}] What is the performance overhead of \method?
    \item[\textbf{Q.4 \method-guided Fuzzing.}] Can \method-based feedback guide test case generation?
    \item[\textbf{Q.5 Configuration Sensitivity Analysis.}] How does \method perform under different configurations?
\end{description}

\emph{Evaluated metamorphic testing methods.} Within the studied methods in \autoref{tab:study}, we chose five metamorphic testing methods whose source code and bug lists are publicly available for evaluation and analysis of the effectiveness of \method based on found bugs. The chosen methods involve 11 metamorphic relations and test three important categories of programs: DBMSs, compilers, and SMT solvers.

For DBMSs, we chose \norec and \tlp. \tlp includes five metamorphic relations for testing the five SQL features: aggregate functions, the queries containing the clauses \lstinline{DISTINCT}, \lstinline{WHERE}, \lstinline{GROUP BY}, and \lstinline{HAVING}. We named them \tlpaggregate, \tlpdistinct, \tlpwhere, \tlpgroupby, and \tlphaving. \norec includes one metamorphic relation for testing query optimizers.

For compilers, we chose \hirgen~\cite{ma2023fuzzing}. We identified two metamorphic relations in \hirgen: \hirgenoptimization and \hirgenmutation, both of which were illustrated in Section 3.3.2 of its paper~\cite{ma2023fuzzing}. 
Although \emi found more bugs than \hirgen, we did not consider \emi, because its source code is not available and its bug reports include minimized bug-inducing test inputs, which we cannot use to evaluate \method.

For SMT solvers, we chose \yinyang~\cite{winterer2020validating} and \sae~\cite{yao2021skeletal}. We identified two metamorphic relations in \yinyang: \yinyangsat and \yinyangunsat, both of which realize \yinyang for testing satisfied and unsatisfied formulas. \sae includes one metamorphic relation.

\emph{Tested programs for evaluated metamorphic testing methods.} Each metamorphic testing method may be applied to more than one target program, for which we chose commonly tested programs. For DBMSs, we chose SQLite and DuckDB, which were tested by \norec and \tlp. SQLite is the most widely deployed DBMS,\footnote{\url{https://www.sqlite.org/mostdeployed.html}} and DuckDB is a successor of SQLite. For compilers, we chose TVM~\cite{chen2018tvm}, which was tested by \hirgen. For SMT solvers, we chose Z3~\cite{de2008z3} and CVC4~\cite{barrett2011cvc4}, which were tested by \yinyang and \sae.  TVM is a popular deep-learning compiler that optimizes the performance of AI modules. Z3 and CVC4 are the two most popular SMT solvers that regularly achieve high ranks in SMT competitions.\footnote{\url{https://smt-comp.github.io}} The chosen programs are written in C/C++, for which we can leverage mature code coverage tools, such as \gcov. 

\emph{Versions of tested programs.} For Q1, we used SQLite (version 3.29.0), DuckDB (commit: a09d2f4 and bc9f086), TVM (commit: 124813f), Z3 (version 4.8.13), and CVC4 (commit: 16c2fe5), which correspond to the major versions in which the bugs from the bug lists were found. For Q2, Q3, and Q5, we used the latest versions supported by these metamorphic testing methods: SQLite (commit c66c77), DuckDB (version 0.5.1), TVM (commit 124813f), Z3 (version 4.13.0), and CVC4 (commit 40eac7f). For Q4, we used the last versions of programs in which the metamorphic testing methods can find bugs: SQLite (commit: 3a461f) and DuckDB (version 0.5.1).

\emph{Seeds for evaluated metamorphic testing methods.} The evaluated metamorphic testing methods require seed inputs from which new test cases are generated. We used the default seeds used in the evaluation of each method. Specifically, \yinyang and \sae adopt the conventional SMT-LIB 2 benchmark as seeds, while other methods implement custom generators to construct suitable seeds, since they require specific input structures. For example, \norec mandates the presence of a \lstinline{WHERE} clause, but does not support queries with \lstinline{GROUP BY}, which makes conventional test suites such as TPC-DS unsuitable for its evaluation.

\emph{Baselines.} We compared \method against line coverage and mutation score. We measured line coverage by \gcov, and mutation score by \mull~\cite{denisov2018mull}.
\mull is a state-of-the-art mutation testing framework that enhances performance by injecting faults into programs during compilation and enabling them respectively through environment variables. We used the 18 default mutators in \mull, including arithmetic and comparison mutations.
\mull identified 19,360, 14,363, 10,396, 6,628, and 21,068 mutations for SQLite, DuckDB, TVM, Z3, and CVC4, respectively. By comparing with them, we gain insights into the benefits of \method against code coverage and mutation score metrics for evaluating metamorphic testing methods.

\emph{Experimental infrastructure.} We conducted all experiments on an AMD EPYC 7763 processor that has 64 physical and 128 logical cores clocked at 2.45GHz. Our test machine uses Ubuntu 22.04 with 512 GB of RAM, and a maximum utilization of 60 cores. We repeated experiments 10 times for statistically significant results.

\subsection*{Q.1 Effectiveness}\label{subsec:Q1}

We investigated whether the bug fixes overlap with the differential code, as examined by \method, and whether \method is correlated with the number of bugs. Overlap indicates that \method covers at least one line of the bug fix. A significant overlap and strong positive correlation would support the applicability of \method in evaluating the bug-finding effectiveness of metamorphic testing methods. We evaluated both based on the historical bugs found by the evaluated metamorphic testing methods.

\emph{Preprocessing test cases.}
We observed that the bug reports of \norec, \tlp, and \yinyang include minimized bug-inducing test inputs, which were reduced to a single bug-inducing input, lacking the second input as well as metamorphic relation, preventing us from directly evaluating \method. \autoref{lst:convertion} shows an example bug found by \norec. Lines 1--4 show the minimized test input, in which only one query exists, while \norec requires a pair of queries to validate the metamorphic relation. To address this in a best-effort manner and apply \method, we manually converted the test cases to trigger the bug using the proposed metamorphic relations. For example, we converted the query in line 4 into a pair of queries in lines 6--7 according to the metamorphic relation of \norec---the first query is a subquery of the second query and both queries should return the same result. Therefore, $t_a$ includes lines 1--4 and 6, and $t_b$ includes lines 1--4 and 7. Such conversion steps were not always possible, due to information lost during minimization. For example, for \norec and \tlp some test cases lacked \lstinline{WHERE} clauses needed for conversion. Based on all found logic bugs, we successfully converted 63\% (27/43) bugs for SQLite and 69\% (20/29) bugs for DuckDB. For \yinyang, we requested 14 original bug-inducing test inputs that include metamorphic relations from the authors. For \hirgen, the metamorphic relations were explicit in all bug-inducing test inputs, so we directly used them. For \sae, most bugs were crash bugs, and we did not find any bug-inducing test inputs for the metamorphic relations. Last, we processed the test inputs only when the buggy behaviors could be reproduced and their corresponding bug fixes could be found in bug reports or were provided by authors.

\begin{figure}
\begin{lstlisting}[caption={Deriving a minimized \href{https://www.sqlite.org/src/tktview?name=a6408d42b9}{test input} into a pair of test inputs that comply to the metamorphic relation.},captionpos=t, label=lst:convertion, escapeinside=@@]
CREATE TABLE t0(c0);
INSERT INTO t0(c0) VALUES (NULL);
CREATE INDEX i0 ON t0(1) WHERE c0 NOT NULL;
SELECT * FROM t0 WHERE (t0.c0 IS FALSE) IS FALSE;
--------------------------------------@$\Downarrow$@--------------------------------------
SELECT COUNT(*) FROM t0 WHERE (t0.c0 IS FALSE) IS FALSE;
SELECT SUM(count) FROM (SELECT ((t0.c0 IS FALSE) IS FALSE) as count FROM t0) as asdf;
\end{lstlisting}

\end{figure}

\begin{table}
    \centering\small
    \caption{Previously found bugs and their relations to \method.}
    \begin{tabular}{@{}lrrr@{}}
\toprule
\textbf{Program} & \textbf{All} & \textbf{Overlapping} & \textbf{Non-overlapping} \\ 
\midrule
SQLite  &  27 & 18 & 9 \\
DuckDB  &  20 & 15 & 5 \\
TVM     &   2 &  2 & 0  \\
Z3      &  14 & 14 & 0  \\
CVC4    &   1 &  1 & 0  \\
\bottomrule
\textbf{Sum:} & 64 & 50 & 14 \\ 
    \end{tabular}
    \label{tab:pre_bugs}

\end{table}

\emph{Bug fix overlap.}
\autoref{tab:pre_bugs} shows the number of bugs and the relations of their fixes to \method. Within the total of 65 bug fixes, 50 bug fixes overlap with \method. 76.9\% of bug fixes are directly located in the code of \method, showing a strong correlation between both. These bugs, whose fixes do not overlap with differential code, occur because buggy functions are called in different code locations.

% https://www.sqlite.org/src/tktview?name=a6408d42b9
% https://www.sqlite.org/src/info/45ff2b1f2693bb02
\begin{figure}

\begin{lstlisting}[caption={SQLite \href{https://www.sqlite.org/src/tktview?name=a6408d42b9}{bug a6408d42}, whose \href{https://www.sqlite.org/src/info/45ff2b1f2693bb02}{fix 45ff2b1f} is overlapped with \method.},captionpos=t, label=lst:example1, escapeinside=&&, linebackgroundcolor={\btLstHL{10, 12-15}}]
CREATE TABLE t0(c0);
INSERT INTO t0(c0) VALUES (NULL);
CREATE INDEX i0 ON t0(1) WHERE c0 NOT NULL;
SELECT COUNT(*) FROM t0 WHERE (t0.c0 IS FALSE) IS FALSE; -- {0}
SELECT SUM(count) FROM (SELECT (t0.c0 IS FALSE) IS FALSE as count FROM t0) as asdf; -- {1}

--- src/expr.c
+++ src/expr.c
@@ -5034,11 +5034,11 @@
 switch( p->op ){
...
  case TK_TRUTH: {
   if( seenNot ) return 0;
   if( p->op2!=TK_IS ) return 0;
-  return exprImpliesNotNull(pParse, p->pLeft, pNN, iTab, seenNot);
+  return exprImpliesNotNull(pParse, p->pLeft, pNN, iTab, 1);
  }
...
\end{lstlisting}

\end{figure}

\emph{An example of the relation overlapping.}
\autoref{lst:example1} shows a bug-inducing test input for a bug found by \norec. The queries in lines 4 and 5 return inconsistent results, which violate \norec. Lines 15 and 16 show the code fix. The gray color represents the code covered by \method. The bug's root cause was an incorrect assumption that $x$ is always not null for the expression \lstinline{(x IS FALSE) IS FALSE}. This assumption is specified by the last parameter of the function \lstinline{exprImpliesNotNull} and is executed for the first query in line 15. The second query evaluates the expression differently and does not execute line 15. The code of \method covers the bug fix in line 15, which is only executed for the first query.

\begin{figure}

\begin{lstlisting}[caption={SQLite \href{https://www.sqlite.org/src/info/6ef984af8972c2eb}{bug 6ef984af}, whose \href{https://www.sqlite.org/src/info/5c118617cf08e17a}{code fix 5c118617} is not overlapped with \method.},captionpos=t, label=lst:example2, escapeinside=&&, linebackgroundcolor={\btLstHL{19}}]
CREATE TABLE t0(c0 TEXT PRIMARY KEY);
INSERT INTO t0(c0) VALUES ('');
SELECT COUNT(*) FROM t0 WHERE (t0.c0, TRUE) > (CAST('' AS REAL), FALSE); -- {0}
SELECT SUM(COUNT) FROM (SELECT ((t0.c0, TRUE) > (CAST('' AS REAL), FALSE)) IS TRUE as count FROM t0) as asdf; -- {1}

--- src/expr.c
+++ src/expr.c
@@ -68,10 +68,13 @@
char sqlite3ExprAffinity(Expr *pExpr){
...
+  if( op==TK_VECTOR ){
+    return sqlite3ExprAffinity(pExpr->x.pList->a[0].pExpr);
+  }
   return pExpr->affExpr;
}
...
aff = sqlite3ExprAffinity(pExpr->pLeft);
...
pCol->affinity = sqlite3ExprAffinity(p);
\end{lstlisting}

\end{figure}

\emph{An example of the relation non-overlapping.}
\autoref{lst:example2} shows another bug-inducing test input for the bug 6ef984af found by \norec. Similarly, \norec finds this bug, because the queries in lines 3 and 4 return inconsistent results. This bug's root cause was a missed case in the function \lstinline{sqlite3ExprAffinity} as fixed by lines 11--13. Although this function is executed for both queries, it is called in different code locations. In line 17, the function call is executed for both queries, and does not trigger the is executed only for the second query, and triggers the special case in line 11, so that the bug is observed.

\begin{figure}
    \centering
    \includegraphics[width=0.6\columnwidth]{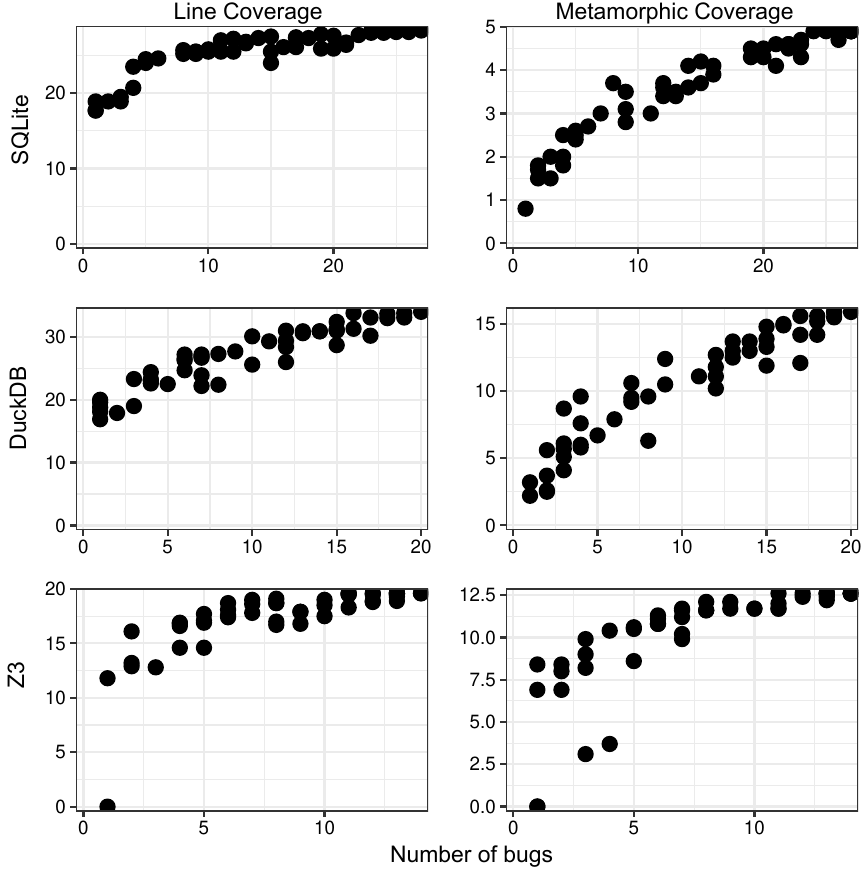}
    \caption{Line coverage and \method against the number of bugs.}
    \label{fig:correlation}
\end{figure}

\begin{table}[]
    \centering\small
    \caption{Pearson Correlation Coefficient of line coverage and \method to bug numbers.}
    \begin{tabular}{@{}lrrr@{}}
    \toprule
    \textbf{Program} & \textbf{Line Coverage} & \textbf{\method} & \textbf{Improvement}  \\
    \midrule
       SQLite   &   0.71    &	0.94 & +0.23 \\
       DuckDB   &   0.86    &	0.94 & +0.08 \\
       Z3       &   0.71    &	0.78 & +0.07 \\
    \bottomrule
                &           & \textbf{Avg:} & +0.13
    \end{tabular}
    \label{tab:correlation}
\end{table}

\emph{Bug correlation.}
Based on the bugs listed in \autoref{tab:pre_bugs}, we randomly sampled varying numbers of bugs to form multiple sets. For each set, we measured line coverage and \method, repeating this process 50 times. \autoref{fig:correlation} illustrates the relationship between the number of bugs in each set and the corresponding line coverage and \method scores. Visually, \method increases more steeply than line coverage as the number of bugs grows. To quantify this observation, we computed the \emph{Pearson Correlation Coefficient} (\pcc)~\cite{cohen2009pearson} between each metric (line coverage and \method) and the number of bugs, as reported in \autoref{tab:correlation}. \pcc measures the linear correlation between two variables, ranging from $-1$ to $1$, with higher values indicating stronger positive correlation. Overall, \method shows a stronger positive correlation with the number of bugs than line coverage as the \pcc of \method is above 0.9 for both SQLite and DuckDB. Compared with line coverage, \method improves the \pcc by an average of 0.13. We excluded TVM and CVC4 from the analysis due to an insufficient number of bugs. 
The mutation score was not included in \autoref{fig:correlation}, as it consistently approaches 100\%—each test case typically triggers buggy behavior, making the metric uninformative in this context.

\result{\method is strongly correlated to the bugs found by metamorphic testing methods as \method overlaps with the fixes of 50 of 64 bugs found by metamorphic testing methods, and its \pcc is above 0.9 for SQLite and DuckDB.}

\subsection*{Q.2 Sensitivity}\label{subsec:Q2}
We evaluated the sensitivity and metric value range of \method for evaluating metamorphic testing methods and compared that with line coverage and mutation score. 
\emph{Sensitivity} refers to the capability of differentiating test inputs and was proposed to evaluate various coverage metrics in fuzzing techniques~\cite{wang2019sensitive}. We used the sensitivity of the various metrics to assess this aspect. Metric value range refers to the possible value range of a metric. In Q1, we showed that differential code is strongly related to bug fixes, which are in the tested program logic. Considering that line coverage subsumes \method, if \method is smaller than line coverage, \method can represent the bug-finding effectiveness in finer granularity. 

\begin{table}
    \centering\small
    \caption{The average number of line coverage, mutation score, and \method of metamorphic testing methods across 10 test suites.}
    \label{tab:cov}
    \begin{tabular}{@{}llrrr@{}}
    \toprule
    \textbf{Program}	& \textbf{Method}	   & \textbf{Line...} & \textbf{Mutation...}	& \textbf{\method} \\
\midrule
\multirow{6}{*}{SQLite}	& \tlpwhere	       		& 21.84\%   & 1.91\%	& 1.96\% \\
 						& \tlpgroupby	   		& 22.49\%   & 1.83\%	& 1.87\% \\
 						& \tlphaving	   		& 22.61\%   & 1.70\%	& 1.42\% \\
 						& \tlpdistinct	   		& 22.14\%   & 1.70\%	& 2.14\% \\
 						& \tlpaggregate	   		& 22.79\%   & 2.47\%	& 2.56\% \\
 						& \norec	       		& 22.68\%   & 1.59\%	& 2.94\% \\
\hline
\multirow{6}{*}{DuckDB}	& \tlpwhere	       		& 20.45\%   & 2.33\%	& 3.37\% \\
 						& \tlpgroupby	   		& 21.47\%   & 2.55\%	& 3.82\% \\
 						& \tlphaving	   		& 21.17\%   & 2.39\%	& 3.45\% \\
 						& \tlpdistinct	   		& 21.29\%   & 2.13\%	& 3.52\% \\
 						& \tlpaggregate	   		& 20.99\%   & 2.64\%	& 4.31\% \\
 						& \norec	       		& 20.35\%   & 1.68\%	& 5.97\% \\
\hline
\multirow{2}{*}{TVM}	& \hirgenmutation	   	& 15.30\%   & 4.36\%	& 3.27\% \\
     					& \hirgenoptimization 	& 17.53\%   &  N/A      & 5.08\% \\
\hline
\multirow{3}{*}{Z3}	    & \yinyangsat	       	& 14.89\%   & 0.01\%	& 2.84\% \\
	    				& \yinyangunsat	       	& 16.49\%   & 0.01\%	& 4.13\% \\
	    				& \sae	       			& 14.80\%   &  N/A   	& 3.75\% \\
\hline
\multirow{3}{*}{CVC4}	& \yinyangsat	       	& 22.89\%   & 0.01\%	& 2.85\% \\
						& \yinyangunsat	       	& 25.21\%   & 0.01\%	& 1.86\% \\
						& \sae	       			& 19.67\%   & N/A       & 2.17\% \\
\bottomrule
    \end{tabular}

\end{table}

\emph{Methodology.}
We measured \method, line coverage, and mutation scores for the test inputs generated by the evaluated metamorphic testing methods.
First, we ran each metamorphic testing method to randomly generate 100 pairs of test inputs as a test suite. 100 is a widely accepted setting for test suite size and has been used for previous work~\cite{andrews2005mutation}. 100 is also a reasonable number, as each coverage data produced by \gcov requires one minute on average on our machine.
We repeated this generation 10 times to generate 10 test suites and measured the average code coverage, mutation score, and \method.
To measure line coverage and \method, we first measured the line coverage for each test input of a test suite, and then calculated the cumulative coverage as the line coverage and the differential coverage as \method for the test suite. 
For mutation score, although \mull already implements various optimizations by injecting all mutations during compilation without the need for recompilation, the time overhead is still significant due to the number of mutants. To measure the mutation scores in a feasible time budget, we ran \mull in 50 parallel threads, which do not depend on each other.
We deemed a mutation to be killed only when the results violate the metamorphic relation. For each testing method, we re-implemented the metamorphic relation within \mull, so that it can identify whether the execution results violate the metamorphic relation. Most testing methods directly produce pairs of inputs, while only \yinyang fuses two inputs into one and examines their consistency, so its testing iteration involves three test inputs. To construct pairs of inputs, we merge the first two test inputs as $t_a$ and deem the third input as $t_b$. 

\emph{Challenges of measuring mutation scores.}
Measuring mutation scores for real-world metamorphic testing methods faces practical issues. \hirgenoptimization and \sae had a high false alarm rate, incurring a 100\% mutation score, which was meaningless, so we ignored their mutation scores. Of the 20 (method, program) configurations, 17 are not affected by this high false alarm rate. \hirgenoptimization detects bugs by checking consistencies before and after optimizations. However, some optimization strategies are incompatible with specific models, and incur errors violating the metamorphic relation. We reported the false alarm issue to the authors of \hirgen, and they confirmed our findings. \sae examines whether a pair of test inputs are satisfied or unsatisfied at the same time, while we observed that executing the second test input may return multiple results, instead of one. We raised GitHub issues to ask the authors of \sae, but have not received any response as of the time of writing this paper. For \yinyangsat and \yinyangunsat on Z3 and CVC4, the mutation score is around 0.01\%. The reason is that \yinyang validates the results only when the first test inputs return the same SAT or UNSAT results. A random mutation can easily cause either input to become invalid or both inputs to return inconsistent results, and \yinyang does not work for such mutated programs.

\begin{table}
    \centering\small
    \caption{Coefficient of variation of line coverage, mutation score, and \method.}
    \begin{tabular}{@{}lrrr@{}}
        \toprule
         \textbf{Program} & \textbf{Line Coverage} & \textbf{Mutation Score} & \textbf{\method} \\
         \midrule
         SQLite & 0.02 & 0.17 & 0.25 \\
         DuckDB & 0.02 & 0.15 & 0.24 \\
         TVM    & 0.10 & N/A  & 0.31 \\
         Z3     & 0.06 & N/A  & 0.19 \\ 
         CVC4   & 0.12 & N/A  & 0.22 \\
         \bottomrule
         \textbf{Avg:} & 0.06 & 0.16 & 0.24 \\
    \end{tabular}
    \label{tab:cv}

\end{table}

\emph{Sensitivity.}
To quantify the sensitivity of various metrics, we evaluated the \emph{Coefficient of Variation} (CV) for \autoref{tab:cov}. CV~\cite{abdi2010coefficient} measures the variability of data independently of the absolute numbers and is defined as the ratio of the standard deviation $\sigma$ to the mean $\mu$, $CV={\frac {\sigma }{\mu }}$. If a metric has a higher value of CV, it is more sensitive than other metrics to distinguish the quality of test suites.

\autoref{tab:cv} shows our results. Across the five programs, the average CV of \method is 0.24, which is 4$\times$ higher than the CV of line coverage, which is 0.06. For each program, the CV of \method is bigger than mutation score, and the line coverage. For mutation scores, we evaluated the average value across SQLite and DuckDB, because mutation scores for other programs are not fully available. The results show that \method is significantly more sensitive to differentiating metamorphic relations than line coverage and also outperforms the mutation score.

\emph{Metric value range.}
\autoref{tab:cov} shows average line coverage, mutation score, and \method across 10 randomly generated test suites by metamorphic testing methods. Overall, \method is 6$\times$ smaller than line coverage. Considering that \method is a subset of line coverage and is strongly correlated to the found bugs, \method can quantify bug-finding capability in finer granularity. The relatively small size of \method suggests that a significant portion of the program remains untested, highlighting the potential for discovering new metamorphic relations to improve testing.

\result{\method can efficiently distinguish the quality of metamorphic testing as it is 4$\times$ more sensitive than line coverage and mutation score for evaluating metamorphic testing. The average value of \method is 6$\times$ smaller than line coverage, while still capturing the tested program logic.}

\subsection*{Q.3 Efficiency}\label{subsec:Q3}
While our major questions are about the evaluation effectiveness of \method, performance overhead is also an important consideration for evaluation metrics, as a high overhead might limit a metric's applicability. We evaluated the execution time of \method, line coverage, and mutation score based on the same experimental set-up and data in Q2. We derived the execution time by multiplying the actual execution time by 50$\times$ as we ran \mull in 50 parallel threads.

\begin{table}
    \centering\small
    \caption{Average time (dd:hh:mm:ss) consumption of measuring line coverage, mutation score, and \method on metamorphic testing methods across 10 test suites.}
    \begin{tabular}{@{}lrrr@{}}
        \toprule
         \textbf{Program} & \textbf{Line Coverage} & \textbf{Mutation Score} & \textbf{\method} \\
         \midrule
         SQLite & 00d:00h:33m:24s & 02d:11h:20m:45s &   00d:00h:36m:01s \\
         DuckDB & 00d:01h:16m:06s & 02d:21h:51m:02s &   00d:01h:22m:27s \\
         TVM    & 00d:04h:06m:44s & 05d:02h:46m:10s &   00d:04h:17m:44s \\
         Z3     & 00d:07h:26m:27s & 11d:21h:09m:55s &   00d:07h:46m:47s \\
         CVC4   & 00d:05h:11m:10s & 240d:01h:31m:40s&   00d:05h:28m:45s \\
         \bottomrule
         \textbf{Avg:} & 00d:03h:18m:22s & 52d:11h:31m:30s& 00d:03h:30m:33s \\
    \end{tabular}
    \label{tab:time}

\end{table}

\emph{Results.}
\autoref{tab:time} shows the average execution time for measuring line coverage, mutation score, and \method across 10 test suites in Q2. Overall, measuring \method consumed the same magnitude of time as measuring line coverage, and 359x less time than measuring mutation score. Mutation score consumes significant time due to its computation complexity. This is particularly noticeable on CVC4, which consumes the most time, because a random mutation can easily cause CVC4 to hang. We set a timeout of 10 minutes for \mull to solve this issue. Compared to the mutation score, \method is more lightweight, because it consumes much less time and does not require test oracles for validating metamorphic relations. Compared to line coverage, \method is a more efficient method, because it can represent the checked code by a metamorphic testing method, but only moderately increases execution time. 

\result{\method is resource-efficient as it requires 359x less time than mutation score and has the same magnitude of time consumption as line coverage.}

\subsection*{Q.4 \method-guided Fuzzing}\label{subsec:Q4}
Multiple fuzzing methods use metrics, such as branch coverage~\cite{afl} and code execution count~\cite{lemieux2018perffuzz}, as guidance to generate or mutate test inputs and have found a large number of bugs.
Apart from evaluating the effectiveness of metamorphic testing methods, we evaluated whether \method can be used as guidance to generate more diverse test inputs increasing the likelihood of finding bugs.

\emph{Methodology.}
For the evaluated metamorphic testing methods, \norec and \tlp involve a test input generation process, while \hirgen, \yinyang, and \sae require user-provided test inputs as $t_a$ for deriving $t_b$, so we evaluated whether \method can help \norec and \tlp generate diverse test inputs. \norec and \tlp are implemented in \sqlancer, which generates test inputs by \emph{Query Plan Guidance} (\qpg)~\cite{qpg}. Given a database, \qpg randomly generates queries and examines their query plans. If no new unique query plan has been seen for a fixed number of iterations, \qpg mutates the database and continues to randomly generate queries on the new database. For a fair comparison, we used \qpg as a reference and reused its test input generation workflow. Specifically, we replaced \qpg as a feedback mechanism with code coverage and \method, and mutated the database if the coverage was not increased for a fixed number of iterations. We reused the instrumentation component of AFL++~\cite{fioraldi2020afl++} to collect branch coverage and derived \method. We refer to both methods as \emph{Code Coverage Guidance} (\ccg) and \emph{Metamorphic Coverage Guidance} (\mcg), respectively.

\emph{Workflow of using \mcg in fuzzing.} At each iteration, the fuzzer generates a candidate test input $t_a$ and applies the metamorphic relation to derive other inputs, such as $t_b$. Both $t_a$ and $t_b$ are executed on the target system with branch-level instrumentation, and $\method(t_a, t_b)$ is computed and accumulated into a global coverage set $\method_{all}$. The fuzzer then generates the next pair of test inputs. If $\method_{all}$ does not grow for a fixed number of iterations, the database is mutated by executing SQL statements to broaden the search space, mirroring the schedule used by \qpg and \ccg.

\emph{Example.} Suppose the fuzzer generates the candidate $t_a:$ \lstinline{SELECT COUNT(*) FROM t WHERE c > 0} and \norec derives $t_b:$ \lstinline{SELECT SUM(c > 0) FROM t}. Executing them on the target database yields $Cov(t_a)$, which covers the branches of the \lstinline{WHERE} clause, and $Cov(t_b)$, which covers the branches of the \lstinline{SUM} aggregate. We then compute $\method(t_a, t_b) = Cov(t_a) \triangle Cov(t_b)$ and accumulate it into $\method_{all}$, after which \norec continues to generate the next pair of inputs. If $\method_{all}$ remains at, for example, $60\%$ and does not increase over 1000 iterations, \mcg mutates the database using SQL statements, such as \lstinline{CREATE INDEX i ON t}, and resumes generating inputs and collecting $\method(t_a, t_b)$. In this way, \mcg steers the fuzzer toward inputs that are more likely to trigger the behavioral discrepancies that metamorphic relations are designed to detect.

\begin{figure}
    \centering
    \includegraphics[width=0.95\columnwidth]{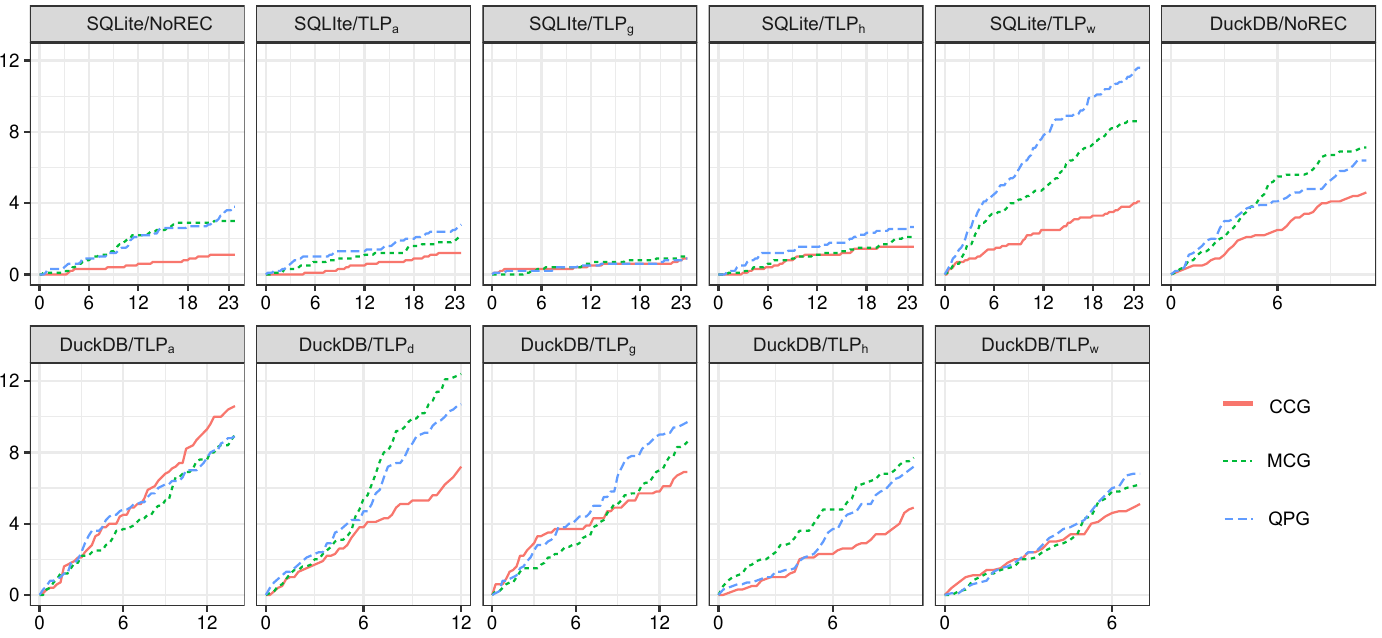}
    \caption{Average number of bugs found by Code Coverage Guidance, Metamorphic Coverage Guidance, and Query Plan Guidance across 10 runs in 24 hours.}
    \label{fig:fuzzing}

\end{figure}

\emph{Results.}
\autoref{fig:fuzzing} shows the average number of bugs found by \ccg, \mcg, and \qpg. DuckDB terminated before 24 hours due to crash bugs. For \tlpaggregate in SQLite, all guidance techniques found more than 100 bug-inducing test cases in five minutes, so it cannot distinguish the contributions of guidance, and we excluded it. Overall, more bugs were found in DuckDB than in SQLite, so the gap between guidance is clearer in DuckDB. On average, \ccg found 4.4 bugs, \mcg found 6.2, and \qpg found 6.5 bugs in 24 hours. \mcg outperforms \ccg on 10 of 11 targets. The result shows that \mcg can help metamorphic testing methods generate more diverse test inputs for finding bugs. The only exception is \tlpaggregate in DuckDB, in which most found bugs are unexpected errors. Unexpected errors refer to unhandled exceptions in programs, such as assertions, and can be found by a test input, instead of a pair of test inputs. \sqlancer can find these unexpected errors by checking error information without \norec and \tlp. Therefore, \ccg helps find more bugs. \mcg outperforms \qpg on 5 of 11 targets. This is unsurprising, as \qpg was designed as a domain-specific feedback metric based on the insight into how DBMSs execute test inputs, while \method can be applied to any metamorphic testing approach, and thus eschews additional domain-specific insights.

\result{Using \method as guidance improved bug-finding efficiency by 41\% on \norec and \tlp compared to code coverage, although domain-specific strategies like \qpg often perform best.}

\subsection*{Q.5 Configuration Sensitivity Analysis}\label{subsec:Q5}

\emph{Various code coverage metrics.}
We evaluated the sensitivity of \method using different kinds of code coverage metrics. As we explained in \autoref{sec:approach}, \method is not restricted to line coverage, so we evaluated whether \method performs similarly under branch and function coverage. Specifically, we reused the same experimental set-up and data in Q2, and examined the CV of \method based on branch and function coverage.

\begin{table}
    \centering\small
    \caption{Coefficient of variation of \method based on branch and function coverage.}
    \begin{tabular}{@{}lrrrr@{}}
        \toprule
            & \multicolumn{2}{c}{\textbf{Branch Coverage}} & \multicolumn{2}{c}{\textbf{Function Coverage}} \\
        \cmidrule(lr){2-3} \cmidrule(lr){4-5}
         \textbf{Program} & \textbf{Branch Coverage} & \textbf{\method} & \textbf{Function Coverage} & \textbf{\method} \\
         \midrule
         SQLite & 0.02 & 0.37 & 0.01 & 0.50 \\
         DuckDB & 0.02 & 0.32 & 0.02 & 0.35 \\
         TVM    & 0.10 & 0.26 & 0.12 & 0.15 \\
         Z3     & 0.19 & 0.20 & 0.07 & 0.22 \\ 
         CVC4   & 0.33 & 0.35 & 0.09 & 0.50 \\
         \bottomrule
         \textbf{Avg:} & 0.13 & 0.30 & 0.06 & 0.34 \\
    \end{tabular}
    \label{tab:sensitivity}
\end{table}

\emph{Results.}
\autoref{tab:sensitivity} shows the CV of \method based on branch and function coverage. On average, \method's CV is 2.3$\times$ more than branch coverage's CV, and 5.7$\times$ more than function coverage. Similar to \autoref{tab:cv}, \method is significantly more sensitive than line, branch, and function coverage to distinguish the quality of metamorphic testing methods. The result shows that \method consistently outperforms code coverage irrespective of at what granularity it is applied.

\emph{Test suite size.}
For Q2, we evaluated \method based on test suites of 100 test inputs. To answer whether different test suite sizes affect the sensitivity of \method, we examined the CV of \method based on test suite sizes of 50 and 200, which are similar to the sizes used in the previous work~\cite{andrews2005mutation}. Due to the resource limitation, we did not consider a bigger size.

\begin{table}
    \centering\small
    \caption{Coefficient of variation of \method based on test suite sizes of 50 and 200.}
    \begin{tabular}{@{}lrrrr@{}}
        \toprule
            & \multicolumn{2}{c}{\textbf{Size 50}} & \multicolumn{2}{c}{\textbf{Size 200}} \\
        \cmidrule(lr){2-3} \cmidrule(lr){4-5}
         \textbf{Program} & \textbf{Line Coverage} & \textbf{\method} & \textbf{Line Coverage} & \textbf{\method} \\
         \midrule
         SQLite & 0.02 & 0.36 & 0.02 & 0.37 \\
         DuckDB & 0.04 & 0.38 & 0.02 & 0.37 \\
         TVM    & 0.15 & 0.56 & 0.11 & 0.29 \\
         Z3     & 0.13 & 0.28 & 0.11 & 0.22 \\
         CVC4   & 0.21 & 0.43 & 0.17 & 0.37 \\
         \bottomrule
         \textbf{Avg:} & 0.11 & 0.40 & 0.09 & 0.32 \\
    \end{tabular}
    \label{tab:size}

\end{table}

\emph{Results.}
\autoref{tab:size} shows the CV of \method based on test suite sizes 50 and 200. We observed that the results are close to the results in \autoref{tab:cv} that \method is around 4$\times$ more sensitive than line coverage. It shows that test suite sizes have no significant effect on the experiments' results. The observation that test suite size has no significant impact on metric evaluation is consistent with previous work~\cite{andrews2005mutation, chen2020revisiting}.

\result{\method performs similarly under branch and function coverage, and under different test suite sizes.}
\section{Discussion}\label{sec:discussion}

\emph{Path to adoption.}
We believe that \method has the potential for wide adoption. From a conceptual perspective, \method is easy to understand, as it calculates the differential code executed by pairs of test inputs. From an implementation perspective, \method is easy to implement as we implemented \method in around 100 lines of Python code for the C/C++ programming language. From an applicability perspective, \method can be implemented on top of existing widely-used coverage measurement tools in other languages, such as JaCoCo for Java, or Coverage.py for Python. From an empirical perspective, our study on widely-used metamorphic relations and systems suggests the effectiveness of \method.

\emph{Improving metamorphic testing methods.}
\method could be used to identify metamorphic relations for code that is not covered by existing metamorphic relations. From our evaluation, we found that most metamorphic testing methods focus on specific features, so their \method is small (\ie, all evaluated metamorphic testing methods have an \method of less than 5\%). For future research on metamorphic testing methods, we believe that systematically designing metamorphic relations to fill \method gaps could significantly improve the bug-finding capability.

\begin{figure}
\begin{lstlisting}[caption={An example of designing new metamorphic relations guided by \method.},captionpos=t, label=lst:design, escapeinside=&&, language=C]
int abs(int x) {
    if (x < 0)
        return -x;
    else if (x == 0)
        return 3; // Bug
    else
        return x;
}
\end{lstlisting}
\end{figure}

\emph{Metamorphic relation design guided by \method.}
Metamorphic relation design is largely a manual and creative task, making it difficult to isolate \method as a quantifiable factor in an automated experiment. Rather, we show an example to show how \method guides the design of new metamorphic relations.
In \autoref{lst:design}, suppose we start with Metamorphic Relation 1 (\emph{MR1}): $abs(x) == abs(-x)$. For example, $abs(3) == abs(-3)$ and $abs(0) == abs(-0)$. \emph{MR1} covers the first and third branches (\ie, lines 2, 3, 6, and 7) differently, so its MC covers the two branches without the second branch at lines 4 and 5. To fill this gap, we need a metamorphic relation in which one test case takes the x == 0 branch while the other takes a different branch. This means the two inputs must include zero and a non-zero value. We can propose Metamorphic Relation 2 (\emph{MR2}): $abs(x) \geq abs(0)$. For instance, $abs(3) \geq abs(0)$ or $abs(-5) \geq abs(0)$. In this case, MC covers all branches, and \emph{MR2} exposes the bug, as $abs(2) < abs(0)$.

\emph{\method on metamorphic relations with multiple inputs and outputs.}
In \autoref{fig:overview}, we defined \method based on metamorphic relations expressed as test pairs $t=(t_a, t_b)$, where $t_a$ and $t_b$ each correspond to a single input. However, some relations require multiple inputs in either $t_a$ or $t_b$. For example, \yinyang, evaluated in \autoref{sec:evaluation}, defines $t_a$ using two inputs. In such cases, we treat the union of the code coverage from all inputs in $t_a$ as the coverage of $t_a$. This step would not affect the effectiveness of \method, as it aims to capture the differential coverage between $t_a$ and $t_b$, reflecting the likelihood of exposing violations of the metamorphic relation.

\emph{Stability of sample sizes.} 
We followed prior work on evaluating coverage metrics~\cite{andrews2005mutation} in choosing test-suite sizes of 50, 100, and 200, and the results above confirm that our conclusions are not specific to a single size. Larger budgets are constrained by a practical bottleneck: \gcov, the official coverage measurement tool that produces the coverage reports consumed by our plugin, generates around 1 GB of disk output per pair of test inputs on large targets such as TVM, which exceeds the disk and I/O capacity of our experimental machines. This bottleneck is specific to the offline measurement pipeline used in Q2, Q3, and Q5; in Q4, we bypass \gcov by computing differential coverage directly in memory using AFL++-style instrumentation, which lets us run \method on millions of test cases over 24-hour fuzzing campaigns. We therefore expect the conclusions from Q2, Q3, and Q5 to hold at larger scales when an in-memory pipeline is used, as Q4 already demonstrates at substantially higher input volumes.

\emph{Threats to Validity.}
The evaluation of \method faces potential threats to validity. A concern is internal validity, that is, the degree to which our results minimize systematic error. \method was compared to other metrics based on randomly generated test suites. The randomness process may limit the reproducibility of our results. To mitigate the risk, we repeated all experiments 10 times to account for potential performance fluctuations.
The other concern is external validity, that is, the degree to which our results can be generalized to and across other metamorphic testing methods, programming languages, and coverage criteria. Regarding the choice of metamorphic relations, we selected the metamorphic relations from our study in \autoref{tab:study}. Typically, one or a few metamorphic relations are presented as a research contribution within a paper; we extracted all 11 MRs from five representative research works, which are widely used in practice (three commonly tested domains), open-source, and include lists of known bugs, which are necessary for our evaluation. According to our study, metamorphic testing methods follow similar testing mechanisms, suggesting that our results generalize to other metamorphic testing methods. Our implementation is based on \gcov, so we focus on C/C++ projects. We expect that evaluating additional programming languages would yield no further insights, as the effectiveness of metamorphic relations depends on the logic of the target system rather than the programming language. Regarding other coverage criteria, in Q.5 of \autoref{sec:evaluation}, we additionally evaluated branch and function coverage to demonstrate generality.

\section{Related Work}\label{sec:related}

\emph{Metamorphic testing.} 
Several studies present a comprehensive overview of metamorphic testing~\cite{segura2018metamorphic, chen2018metamorphic, segura2016survey}. As shown in \autoref{tab:study}, we investigated the evaluation metrics of ten metamorphic testing methods, which have cumulatively found thousands of bugs. 
These methods have implicitly relied on the intuition behind \method when designing metamorphic relations for bug detection. For example, \emi~\cite{le2014compiler} describes its goal as ``exploiting the interplay between dynamically executing and statically compiling'', which effectively encourages behavioral differences across executions. Such works, however, treat this idea only as an informal guiding principle for constructing metamorphic relations, rather than as a measurable quantity.
In this work, rather than studying existing metamorphic relations or proposing a new one, our core contribution is a novel coverage metric to evaluate metamorphic testing methods and a comprehensive evaluation thereof.

\emph{Evaluating metamorphic relations.}
Several empirical studies have investigated what makes a metamorphic relation effective, as summarized in Table~\ref{tab:related_work_comparison}. Chen~\etal~\cite{chen2004case} performed case studies on shortest-path and critical-path implementations, hypothesizing that metamorphic relations whose initial and follow-up executions diverge more are more likely to expose faults; however, they left ``difference between two executions'' undefined as a computable quantity, so the guidance remains qualitative. Mayer~\etal~\cite{mayer2006empirical} reached a complementary qualitative conclusion on small numerical programs, observing that metamorphic relations with richer semantic content tend to be stronger. Asrafi~\etal~\cite{asrafi2011testing} correlated the \emph{cumulative} line and branch coverage of source and follow-up test cases with fault-detection effectiveness on two small programs (TCAS, 173~LOC; KNAPSACK, 780~LOC). Cumulative coverage, however, takes the \emph{union} $Cov(t_a) \cup Cov(t_b)$ rather than the symmetric difference, and therefore cannot distinguish metamorphic relations that exercise the same code with both inputs from those that exercise disjoint paths, which corresponds to the situation in our motivating example (Section~\ref{lst:intro}).

The closest prior work is Cao~\etal~\cite{cao2013correlation}, who formally defined six execution-distance metrics, including the Branch Coverage Manhattan Distance~(BCMD), which counts the number of branches covered by exactly one of $t_a$ and $t_b$, and evaluated the correlation between BCMD and fault-detection effectiveness across seven packages. Their metrics and \method are conceptually related: under branch coverage, BCMD equals $|\method(t_a, t_b)|$, the \emph{cardinality} of the differential set. \method retains the set itself rather than collapsing it to a scalar, making it \emph{linkable to code}: each element points to a specific line or branch, so we can verify that the metric overlaps with bug-fix locations~(Section~\ref{sec:evaluation}, Q1) and identify untested regions to guide the design of new metamorphic relations---neither of which a scalar score supports.
Overall, all prior studies validate their metrics on manually crafted metamorphic relations and small programs with seeded mutations; only Cao~\etal~\cite{cao2013correlation} include a single larger subject (bash). In contrast, we evaluate \method on five widely used, production-grade metamorphic testing methods (\norec, \tlp, \hirgen, \yinyang, \sae) applied to industrial systems (SQLite, DuckDB, TVM, Z3, CVC4), and correlate it with 64 real-world bugs. Furthermore, none of the prior metrics has been used to guide test-case generation in fuzzing; we show that \method can serve as a coverage signal for feedback-driven fuzzing, finding 41\% more bugs than branch-coverage guidance on \norec and \tlp.

\begin{table*}[t]
\centering
\small
\caption{Comparison of \method with related empirical studies on evaluating metamorphic relations.}
\label{tab:related_work_comparison}
\renewcommand{\arraystretch}{1.25}
\begin{adjustbox}{max width=\textwidth}
\begin{tabular}{p{3.0cm} p{2.2cm} p{2.4cm} p{2.4cm} p{2.6cm} p{3.6cm}}
\toprule
\textbf{Dimension} & Chen~\etal~\cite{chen2004case} & Mayer~\etal~\cite{mayer2006empirical} & Asrafi~\etal~\cite{asrafi2011testing} & Cao~\etal~\cite{cao2013correlation} & \textbf{\method (ours)} \\
\midrule
Formally defined metric & \xmark\ & \xmark\ & \xmark\ & \cmark\ (distance metrics) & \cmark\ (metamorphic coverage) \\
Link to specific code & \xmark & \xmark & \xmark & \xmark\ & \cmark\ (per-line / per-branch) \\
Fuzzing guidance & \xmark & \xmark & \xmark & \xmark\ & \cmark\ (41\%$\uparrow$ bugs found) \\
Metamorphic relations & Author-crafted & Author-crafted & Author-crafted & Author-crafted & 11 real-world relations \\
Subject programs & 2 graph programs \newline ($\le$\,$\sim$200~LOC) & 6 packages \newline ($\le$\,$\sim$900~LOC) & 2 programs \newline ($\le$\,780~LOC) & 7 packages \newline (95--59{,}846~LOC) & 5 industrial systems \newline (15{,}543--51{,}974~LOC) \\
Bugs analyzed & Seeded mutants & Seeded mutants & Seeded mutants & Seeded + real faults & 64 real-world bugs \\
\bottomrule
\end{tabular}
\end{adjustbox}
\end{table*}

\emph{Oracle coverage.}
Some metrics have been proposed to evaluate test oracles, mostly assertions, which typically check invariants when executing a single test input. Koster \etal~\cite{koster2007state} proposed state coverage, which measures the percentage of the output-relevant variables checked by an assertion. Vanoverberghe \etal~\cite{vanoverberghe2012state} improved state coverage to measure the percentage of only program variables that are read by the assertion. Schuler \etal~\cite{ schuler2013checked} proposed checked coverage, which measures the percentage of statements checked by an assertion based on all statements that influence the assertion by control flow or data flow. Hossain \etal~\cite{hossain2023measuring} proposed to reduce the total space of checked coverage by measuring only the statements that influence at least one value checked by the assertion. In this work, we propose \method to evaluate metamorphic testing, which compares discrepancies between two test inputs instead of an input.

\emph{Coverage criteria.}
A large body of work considers the relationship between coverage criteria and fault detection. Gligoric \etal~\cite{gligoric2013comparing} examined the correlations of various coverage to mutation testing and concluded that branch coverage and intra-procedural acyclic path coverage perform best to predict the mutation score, which is assumed to be a silver criterion for evaluating test suites. Gopinath \etal~\cite{gopinath2014code} further studied the problem and concluded that statement coverage performs best. Inozemtseva \etal~\cite{inozemtseva2014coverage} investigated the correlation between various coverage criteria and the mutation score for different random subsets of test suites and found a low to moderate correlation between coverage and effectiveness when the number of test inputs in the suite is controlled for. Kakarla~\cite{kakarla2010analysis} and Inozemtseva~\cite{inozemtseva2012predicting} demonstrated a linear relationship between mutation score and various coverage criteria for individual programs. In contrast, in this work, we propose a novel coverage criterion \method for evaluating metamorphic testing.

\section{Conclusion}

In this paper, we have proposed a novel coverage metric, \emph{Metamorphic Coverage} (\method), to evaluate the quality of metamorphic testing methods. The core idea of \method is that a bug can be observed if the faulty code path is executed by a test input, but not the other test input. Therefore, if a metamorphic testing method covers more differential code between the execution of pairs of test inputs, it is more likely to find bugs. The results show that \method is strongly correlated to the bugs found by metamorphic testing methods as the code covered by \method overlaps with the fixes of 50 of 64 bugs found by the five metamorphic testing methods. \method is 4$\times$ more sensitive than line coverage in distinguishing the quality of metamorphic testing methods and is similarly lightweight in terms of time consumption as line coverage. Despite these promising results, \method{} is no panacea, similar to code coverage, as achieving a high \method is possible even if the metamorphic test oracle has low bug-finding effectiveness and \emph{vice versa}. In the future, we believe that \method can be broadly applied to assess metamorphic testing methods and improve test-case generation by using \method as a metric in feedback-guided automated testing.

\section{Data Availability}
The artifact is available at \url{https://github.com/nus-test/metamorphic_coverage}.

\begin{acks}
This research is supported by the National Research Foundation, Singapore, and Cyber Security Agency of Singapore under its National Cybersecurity R\&D Programme (Fuzz Testing <NRF-NCR25-Fuzz-0001>). Any opinions, findings and conclusions, or recommendations expressed in this material are those of the author(s) and do not reflect the views of National Research Foundation, Singapore, and Cyber Security Agency of Singapore.
\end{acks}

\bibliographystyle{plain}
\bibliography{references}

\end{document}